\documentclass[preprint,
 superscriptaddress,%groupedaddress,%unsortedaddress,%runinaddress,%frontmatterverbose, %preprint,%showpacs,preprintnumbers,
 nofootinbib,%nobibnotes,
 bibnotes,
 amsmath,amssymb, aps,%pra,%prb,
 prd,
floatfix,
]{revtex4-1}%%%%%%%%%%%%%%%%%%%%%%%%%%%%%%%%%%%%%%%%%%%%%%%%%

\usepackage{graphicx}
\usepackage{epstopdf}
\usepackage{bm}
\usepackage{bbm}

\usepackage[colorlinks]{hyperref}
\hypersetup{
     colorlinks   = true,
     citecolor    = red,
 linkcolor=blue
}
\usepackage{color}
\usepackage{soul}
\usepackage{amsmath}
\usepackage{amssymb}
\usepackage{mathtools}
\usepackage{xfrac}
\usepackage{ytableau}
\usepackage{rotating}
\usepackage{scrextend}

\newcommand{\be}{\begin{eqnarray}}
\newcommand{\ee}{\end{eqnarray}}
\newcommand{\la}{\lambda}
\newcommand{\lt}{\widetilde{\lambda}}

\newcommand{\ep}{\epsilon}

\newcommand{\tY}{\widetilde{Y}}
\newcommand{\nt}{\widetilde{n}}

\newcommand{\ad}{{\dot{a}}}
\newcommand{\al}{{a}}

\newcommand*\oline[1]{%
   \vbox{%
     \hrule height 0.5pt%                  % Line above with certain width
     \kern0.5ex%                          % Distance between line and content
     \hbox{%
       \kern-0.0em%                        % Distance between content and left side of box, negative values for lines shorter than content
       \ifmmode#1\else\ensuremath{#1}\fi%  % The content, typeset in dependence of mode
       \kern-0.0em%                        % Distance between content and left side of box, negative values for lines shorter than content
     }% end of hbox
   }% end of vbox
}

\begin{document}

\graphicspath{{Figures/}}

\title{
Constructing effective field theories via their harmonics
}
\preprint{IPMU19-0001}

\author{Brian Henning}
\affiliation{Department of Physics, Yale University, New Haven, Connecticut 06511, USA}
\affiliation{D\'epartment de Physique Th\'eorique, Universit\'e de Gen\`eve, 24 quai Ernest-Ansermet, 1211 Gen\`eve 4, Switzerland}

\author{Tom Melia}
\affiliation{Kavli Institute for the Physics and Mathematics of the Universe (WPI),The University of Tokyo Institutes for Advanced Study, The University of Tokyo, Kashiwa, Chiba 277-8583, Japan}

\begin{abstract}
We consider the construction of  operator bases for massless, relativistic quantum field theories, and show this is equivalent 
to obtaining the harmonic modes of a physical manifold (the kinematic Grassmannian), upon which observables have support. This enables
us to recast the approach of effective field theory (EFT) through the lens of harmonic analysis. We explicitly construct harmonics  corresponding to low 
mass dimension EFT operators.
\end{abstract}

\date{\today}

\maketitle
\section{Introduction}

The approach of EFT is to consider all possible contributions to a given physical observable.
Particle scatterings and decays  only have support on a physical manifold where momentum conservation
and on-shell conditions are satisfied. These constraints manifest as what are termed equations of motion (EOM) and integration by parts (IBP) relations
between operators in the EFT, and have been the subject of extensive study spanning the past few 
decades~\cite{Buchmuller:1985jz,Grzadkowski:2010es,Jenkins:2013zja,Lehman:2014jma,Lehman:2015via,
Henning:2015daa,Henning:2015alf,Henning:2017fpj}.

In a series of papers~\cite{Henning:2015daa,Henning:2015alf,Henning:2017fpj}, it was shown that these 
constraints are ultimately a consequence of
the  Poincar\'e symmetry of spacetime; this insight enabled a systematic enumeration of basis elements ({\it i.e.}
operator counting) in an EFT. In particular, by considering a larger spacetime symmetry---that of the 
conformal group---it was shown the operator basis naturally consists of conformal primary operators, which
could then be counted using Hilbert series techniques. 

In this note, we put operator construction on the same footing as 
operator enumeration, by detailing the 
systematic construction of the conformal primary operators that  provide a privileged choice of 
basis for the $S$-matrix of the theory (for other approaches to operator basis construction, see~\cite{Henning:2015daa,Lehman:2015via,Lehman:2015coa,Henning:2017fpj,Gripaios:2018zrz,Shadmi:2018xan,Criado:2019ugp}). 
The presentation is designed to accompany the paper~\cite{letter}, which considers more generally the entire operator spectrum  (not just Lorentz scalars), as is relevant for more general correlation functions. This note also 
 proceeds more heuristically than~\cite{letter}---in particular, by including a number of worked examples---and omits many mathematical
details. We have endeavoured to provide pointers to~\cite{letter} in the relevant places. We would, however, like to refer 
the interested reader to \cite{letter} for a reinforced connection to ideas in conformal field theory (CFT), 
and modern (Hamiltonian truncation) non-perturbative methods.

We consider four dimensional relativistic theories of massless particles, and allow for all particle spins by working with  
spinor helicity variables, which encode both kinematic and helicity information.
 In these variables a $U(N)$ action on the phase space of $N$ particles is revealed, which generalises
the $U(1)^N\subset U(N)$ particle little group scalings. This symmetry plays a crucial role, first via a duality with the conformal group $SU(2,2)\simeq SO(4,2)$ that in~\cite{letter} we term `conformal-helicity duality', and second via its 
symmetry breaking pattern which, in the case of EFTs, is down to $U(N-2)\times U(2)$, identifying the physical manifold as the  Grassmann manifold $G_2(\mathbb{C}^N) = U(N)/U(N-2)\times U(2)$ (the kinematic Grassmannian~\cite{ArkaniHamed:2009dn}).

A new picture of EFT emerges---that of harmonic analysis on the Grassmann manifold. There is a tight
analogy with the harmonic analysis of a sphere: functions $f=f(x,y,z)$,  with coordinates  subject
to the constraint $x^2+y^2+z^2=1$, can be expanded in terms of spherical harmonics on the sphere,
$f=\sum_{l,m}c_{lm}Y_{lm}$. In the EFT case, observables ${O}(\{ p_i\})$, subject to the constraints $p_i^2=0$
 and $\sum_i p_i^\mu=0$ involving particles of any spin can be similarly decomposed into harmonics of 
the Grassmannian, ${O}=\sum_{\vec{l}}c_{\vec{l}}\tY_{\vec{l}}$ (with Wilson coefficients $c_{\vec{l}}$, and
 with $\vec l$ a vector of quantum numbers to be specified below). 
 For the case of the sphere, harmonic polynomials in $x$, $y$ and $z$ are annihilated by the Laplacian, $\nabla^2$; these 
 form a basis of spherical harmonics when restricted to the sphere. For the EFT case, we will construct harmonic polynomials in spinor variables which are annihilated by a 
 generalised Laplacian operator, $K$, that turns out to be the special conformal generator (whence 
 $\tY_{\vec{l}}$ are primary); these form a basis for the $S$-matrix. 
 
 The note has the following structure. In Sec.~\ref{sec:construct} we detail the construction of the EFT harmonics, presenting
 the main result from~\cite{letter} and providing additional heuristic motivation. In 
 Sec.~\ref{sec:lowmass} we use this result to explicitly construct low-lying harmonics thereby providing EFT bases at low mass dimension.
Sec.~\ref{sec:discuss} concludes.

\section{Constructing EFT harmonics}
\label{sec:construct}

EFT quantifies all possible $S$-matrix elements between an $|in\rangle$ state in a multi-particle Fock space and the vacuum,
\be
\langle 0 | S |in\rangle \,.
\ee
We consider  massless asymptotic particle states\footnote{Massive states can be
described via two massless states (up to an $SU(2)$ little group redundancy), 
see {\it e.g.}~\cite{Arkani-Hamed:2017jhn}; we will leave extensions in this direction to future work.} labelled by kinematic (three momenta), helicity, and possibly some internal quantum numbers. Moreover, we consider multi-particle 
states that are built from distinguishable particles, deferring a discussion 
on exchange symmetry to Sec.~\ref{sec:discuss}.

We encode the kinematic information using spinor helicity variables,
\be
p^\mu_i (\overline{\sigma}^{ \ad a})_\mu =  \lt^{\ad}_i \lambda^{i \,a } \,, ~~~ \lt_i^{\ad}=(\la^{i\,a})^*\,,
\ee
with $\nobreak{a, \ad=1,2}$ the usual Lorentz indices and $\nobreak{i=1,\ldots,N}$ a particle, or flavour, index (raised on $\la$ and lowered on $\lt$ to anticipate
the action of a $U(N)$ symmetry),
such that $S$-matrix elements
\be
&&\langle 0 | S | \la^1,\lt_1, h_1; \ldots ; \la^N,\lt_N,h_N \rangle = f(\{ \la^i,\lt_i\}) \,\delta^{(4)}(\sum_{i=1}^N  \lt^{\ad}_i \lambda^{i \,a } )  \,,
\label{eq:ffunc}
\ee
where $f(\{ \la^i,\lt_i\})$ is a Lorentz scalar function of the spinor variables. In eq.~\eqref{eq:ffunc} we labelled states in the Fock space with spinors to encode the kinematic information, and with helicities $h_i$. In these variables,
Lorentz invariant phase space  is written as,
\be
d\Phi_N = \prod_{i=1}^N d^4 p_i \delta^+(p_i^2)= \prod_{i=1}^N \frac{d^2\la^i \,d^2\lt_i}{\text{Vol}(U(1))}\,,
\label{eq:phsp1}
\ee
where $\delta^+(p_i^2) = \delta(p_i^2)\theta(p^0)$ and the volume of the little group $\text{Vol}(U(1))=2\pi$.
 
 We are interested  in a basis for the functions $f$ in eq.~\eqref{eq:ffunc}. 
 That this is equivalent
 to constructing an EFT basis, taking into account EOM and IBP, follows from the standard introduction
 of local operators as products of interpolating fields---see~\cite{Henning:2017fpj} for a detailed
 discussion on this point. Note that in using spinors, we automatically take into account the EOM ({\it i.e.} the momenta are on-shell). The fields are required to transform under Poincar\'e 
 in the way dictated by the helicity of the asymptotic state. For example, $\la^{i\,a}$ transforms in the $(j_1,j_2)=(\frac{1}{2},0)$
 representation (rep) of Poincar\'e, and thus interpolates a negative helicity fermion $\psi_L$; $\lt^\ad_i\lt^{\dot{b}}_i$ transforms in  the $(0,1)$ rep and interpolates a positive helicity massless spin-1 state, (the field-strength operator $F_R=\frac{1}{2}(F+i \widetilde{F})$ in spinor variables);  pairs of $\lt^\ad_i \la^{i\,a}$ imply a derivative acting on the interpolating fields in the operator. In other words,
 \be
 F_R^{\ad \dot{b}} = \int\frac{d^2\lambda d^2\lt}{\text{Vol}(U(1))} \left( \lt^{\ad} \lt^{\dot{b}} e^{\frac{i}{2}\la^a\lt^\ad x_{a\ad}}a^\dagger + \text{h.c.} \right) \,,
 \ee
 {\it etc.}
  In this way, $f$  transforms under the asymptotic particle little groups with the correct helicity weight.

The delta function in eq.~\eqref{eq:ffunc} enforces total momentum conservation,
\be
P^{ \ad\al} = \sum_{i=1}^N  \lt^{\ad}_i \lambda^{i \,a }  =0 \,.
\label{eq:momcon}
\ee
This equation is a constraint on the variables $\la$ that fixes the functions $f$ in eq.~\eqref{eq:ffunc} to lie on
some manifold $\subset \mathbb{C}^{2N}$. This manifold is well known in the literature to be the Grassmannian, $G_2(\mathbb{C}^N)$~\cite{ArkaniHamed:2009dn}.
 Fixing a Lorentz frame and writing, 
\be
\left( \begin{array}{c c c c} 
\lambda^{1\,1} &  \lambda^{2\,1} & \cdots & \lambda^{N\,1} \\ 
\lambda^{1\,2} &  \lambda^{2\,2} & \cdots & \lambda^{N\,2}
\end{array}\right) = 
\left( \begin{array}{c } 
{\bf u} \\
{\bf v}
\end{array}\right) \,,
\label{eq:uv}
\ee
one sees that the vectors ${\bf u}$ and ${\bf v}$ define a 2-plane; under Lorentz transformations  ${\bf u}$ and ${\bf v}$ rotate within the plane so that, modulo these transformations, Lorentz invariant phase space is described as the set of 2-planes
that intersect the origin in in $\mathbb{C}^{2N}$, which defines $G_2(\mathbb{C}^N)$. A more general manifold is obtained if one does not mod out by Lorentz rotations~\cite{letter}---this case
is most easily analysed by considering the breaking of the $U(N)$ symmetry (under which ${\bf u}$ and ${\bf v}$ transform
as fundamentals) by eq.~\eqref{eq:momcon} down to $U(N-2)$. The manifold that the coset $U(N)/U(N-2)$ defines is known as 
the Stiefel manifold. (In the present case, the coset construction of the Grassmannian is $U(N)/U(N-2)\times U(2)$, with the 
extra $U(2)$ being the Lorentz transformations and a complex phase that are further modded out.)

Returning to the analogy with the sphere where the Laplacian $\nabla^2$ in essence forms an adjoint to $|{\bf r}|^2$, we construct  the adjoint operator to $P^{\ad\al}$ as
\be
K_{\ad\al} = -\sum_{i=1}^N \frac{\partial}{\partial \lt_i^{\ad}} \frac{\partial}{\partial \la^{i \,\al}} \,,
\ee
which is the generator of special conformal transformations in spinor variables. The harmonic modes of the Grassmannian manifold are those annihilated by $K$; they are thus identified with primary conformal operators. To construct
a basis for the functions $f$ in eq.~\eqref{eq:ffunc} we therefore turn to constructing such harmonic polynomials.

\subsection{Harmonics from  Young diagrams}

Let us build basis polynomials out of $n$ $\la$s and $\nt$ $\lt$s, at fixed $N\ge2$. Because the polynomials are Lorentz scalars,
 $n$ and $\nt$ must be even, with the spinors contracted as
\be
[j_1 j_2]\ldots[j_{\nt-1} j_{\nt}] \,\langle i_1i_2\rangle\ldots\langle i_{n-1} i_n \rangle,
\label{eq:halfraised}
\ee
where we use bracket notation $\langle i j \rangle=\lambda^{i\,\al}\lambda^j_\al$, $[ i j]=\lt_{i\,\ad}\lt_j^\ad$, and where the indices $i_1..i_n,j_1..j_{\nt}$ are (unspecified as yet) particle indices. 

We consider raised particle number indices on $\la$ as $U(N)$ indices, 
such that $\la^{i\,\al}$ transforms
under $SL(2,\mathbb{C})\times U(N)$ as spinor $\times$ fundamental. Similarly, $\lt^{\ad}_i$ transforms as (the complex conjugate representation) spinor $\times$ anti-fundamental. That is, the indices $i_1$ to $i_n$ in eq.~\eqref{eq:halfraised} can be interpreted as (raised) $U(N)$ indices, and the indices $j_1$ to $j_{\nt}$ can be interpreted as (lowered) conjugate $U(N)$ indices. The latter can be raised using the  epsilon tensor, 
\be
[j_1 j_2]\ep^{j_1j_2 k_1 .. k_{N-2}}\ldots [j_{\nt-1} j_{\nt}]\ep^{j_{\nt-1}j_{\nt} l_1 .. l_{N-2}} \,\langle i_1i_2\rangle\ldots\langle i_{n-1} i_n \rangle,
\label{eq:allraised}
\ee
with summation over all $j$ indices.

\begin{figure}
\includegraphics[width=5cm]{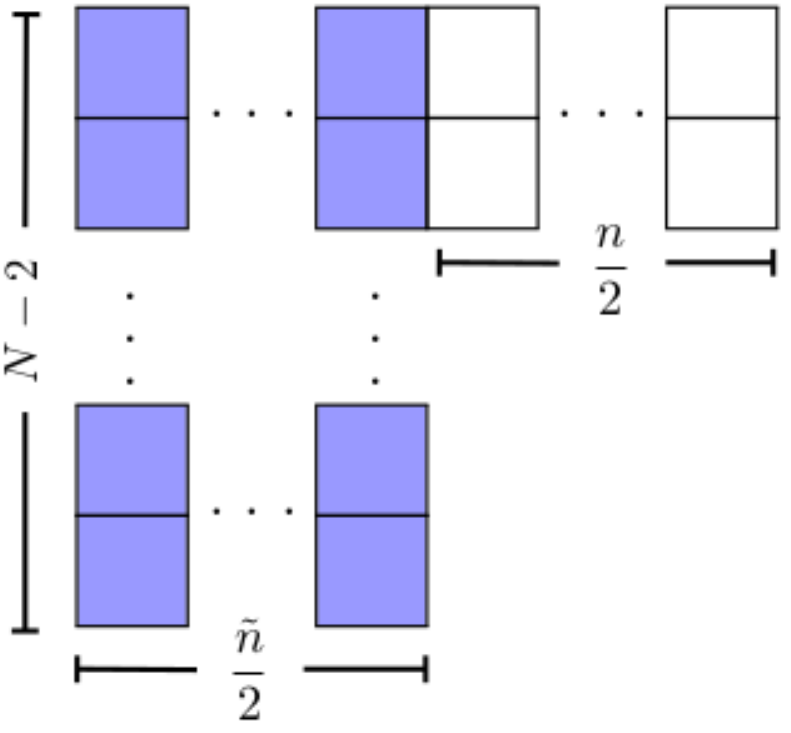}
\caption{
 Young diagram for the harmonic modes of the Grassmannian.}
 \label{fig:glue}
 \end{figure}
The key result of~\cite{letter} is that the basis polynomials furnish a particular representation of $U(N)$, labelled by eigenvalues $n$ and $\nt$. Equivalently one can label by mass dimension $\Delta$, and helicity $h$,\footnote{To provide a translation to the notation used in~\cite{letter}, here $n=l_1+l_2$ and $\nt=\tilde{l}_1+\tilde{l}_2$ in the Lorentz scalar case where $l_1=l_2$ and $\tilde{l}_1=\tilde{l_2}$.
We note that more general non-Lorentz-scalar operators are further labelled by spin eigenvalues, $j_1$ and $j_2$.}
\be
\Delta &=& \frac{1}{2}(n+\nt)+N \,,\\
h&=&\frac{1}{2}(n-\nt).
\ee

 Finite dimensional representations of $U(N)$ are in one-to-one correspondence with Young diagrams---see {\it e.g.}~\cite{Georgi:1982jb}. That is, the Young diagrams encode the symmetrisation pattern to be applied to the indices in eq.~\eqref{eq:allraised}, to form a $U(N)$ irreducible representation. The particular Young diagram that renders eq.~\eqref{eq:allraised} a harmonic mode
 of the Grassmannian is given in Fig.~\ref{fig:glue}. The indices $k_1\ldots k_{N-2}$ in eq.~\eqref{eq:allraised} are associated
 with the first column which is shaded blue (to indicate it corresponds to $\lt$ indices raised with an epsilon tensor); the indices $l_1\ldots l_{N-2}$  in eq.~\eqref{eq:allraised} are associate with the right-most blue column; the indices $i_1,i_2$ with the left-most
 unshaded column; and, the indices $i_{n-1},i_n$ with the final column:
 \be
 \includegraphics[width=4cm]{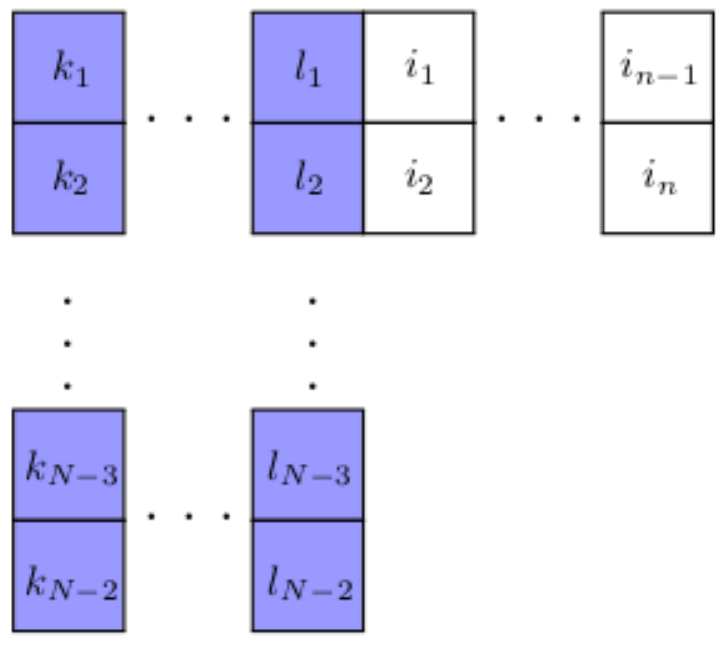} \,. \nonumber 
 \ee
 
A basis for the $U(N)$ rep is supplied by semi-standard Young tableaux, as discussed in the next subsection. For now, we want to reflect upon why it is that this representation is primary.

To begin to understand this result, let us start by considering holomorphic operators---that is,
functions consisting purely of $\la$s. These are obviously primary (annihilated by $K$). We consider basis functions that are polynomials in a fixed number 
$n$ of $\la$s. These $\la$ carry two indices, $\la^{i\,\al}$. A simple but important observation is that if a symmeterisation pattern is applied to one index, the other index automatically inherits this pattern. For example, 
\be
\lambda^{i\,\al} \lambda^{j\,b} + (i\leftrightarrow j) = \lambda^{i\,\al} \lambda^{j\,b}+\lambda^{j\,\al} \lambda^{i\,b} \,,
\ee
is a symmeterisation in particle indices $i$ and $j$, but the resulting expression is also symmetric in $\al$ and $b$. Similarly,
\be
\lambda^{i\,\al} \lambda^{j\,b} - (i\leftrightarrow j) = \lambda^{i\,\al} \lambda^{j\,b}- \lambda^{j\,\al} \lambda^{i\,b} \,,
\ee
anti-symmeterises in $i$ and $j$; the anti-symmetery is inherited by $\al$ and $b$ as well. This works for general 
symmeterisation patterns that are encoded by the Young diagrams. So, when
a polynomial in $n$ $\la$s is organised into a singlet representation of $SL(2,\mathbb{C})$---corresponding to a Young
diagram with $n/2$ boxes in the first row and $n/2$ boxes in the second row---the $U(N)$ indices inherit the exact same symmeterisation pattern,
\be
f^{(n)}_{\text{hol}}(\{\la\}) &=& g_{SL(2,\mathbb{C})}\otimes g_{U(N)}  \nonumber \\
&=& \begin{array}{c}  \includegraphics[width=5cm]{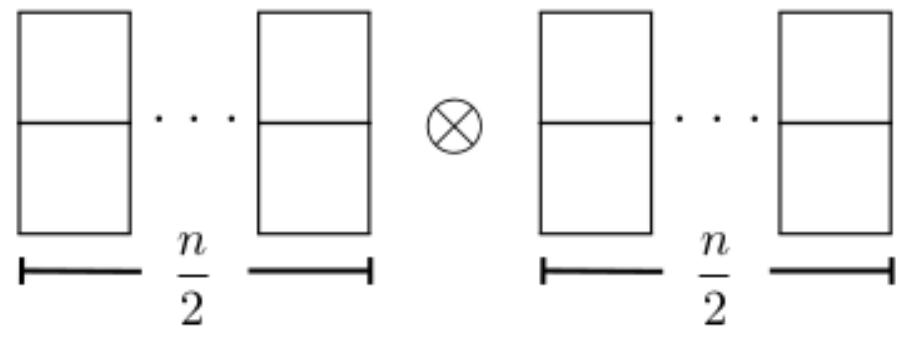} \end{array} \,,
\ee
Note that this implies that $U(N)$ representations corresponding to Young diagrams with more than two rows---{\it i.e.} that are anti-symmetrised on more than two indices---can
never be constructed, {\it e.g.} $\lambda^{i\,\al} \lambda^{j\,b}\lambda^{k\,c} + (\text{anti-sym in } i,j,k) =0$, for all $a$, $b$, $c$.

The above considerations apply to anti-holomorphic basis functions in $\nt$ $\lt$s: again, the $U(N)$ representation
is dictated by the symmeterisation pattern on the Lorentz indices such that the functions are Lorentz scalars,
\be
f^{(\nt)}_{\text{anti-hol}}(\{\lt\}) &=& g^\star_{SL(2,\mathbb{C})}\otimes \overline{g}_{U(N)}  \nonumber \\
&=& \begin{array}{c}  \includegraphics[width=5cm]{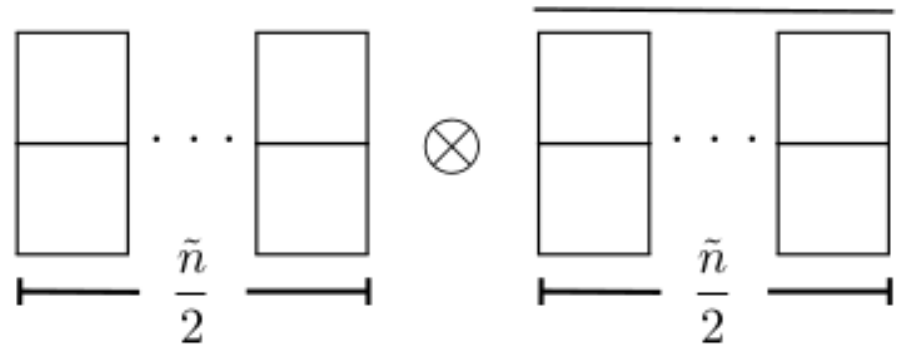} \end{array} \nonumber \\
&=& \begin{array}{c}  \includegraphics[width=5cm]{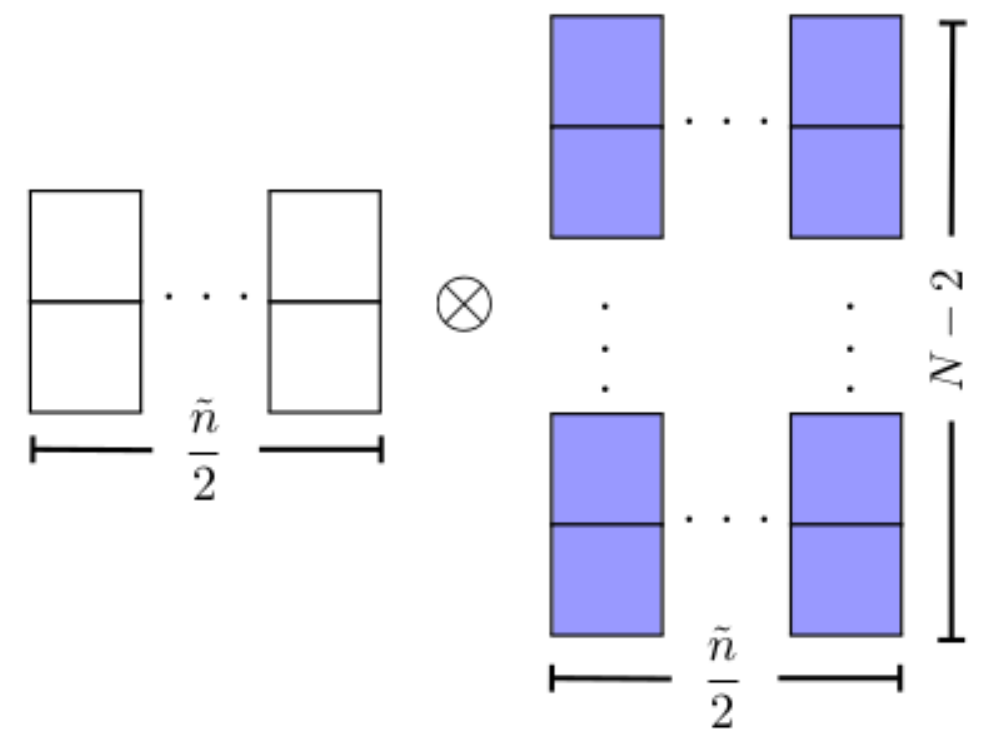} \end{array} \,,
\ee
where we used a barred Young diagram to denote the conjugate $U(N)$ representation;  in the last equality we redrew this as the $\ep$ tensor conjugated diagram.

Now we turn to the non-holomorphic case, concerning $n$ $\la$s and $\nt$ $\lt$s. Such operators only appear for $N\ge 4$, which reflects the familiar fact that Mandelstam invariants are trivial for $N\le 3$~\cite{letter}. The $\la$s and $\lt$s  separately
have their $SL(2,\mathbb{C})$ indices symmeterised into the Lorentz scalar patterns as in the holomorphic and
anti-holomorphic cases above; again the $U(N)$ indices and conjugate $U(N)$ indices will inherit the same pattern.
What is different this time, is that now the resulting $U(N)$ representation is reducible,
\be
f^{(n,\nt)}(\{\la,\lt\}) &=&  g_{SL(2,\mathbb{C})}\otimes g^\star_{SL(2,\mathbb{C})}\otimes ( \overline{g}_{U(N)} \otimes g_{U(N)} ) \nonumber \\
&=&\begin{array}{c}  \includegraphics[width=10cm]{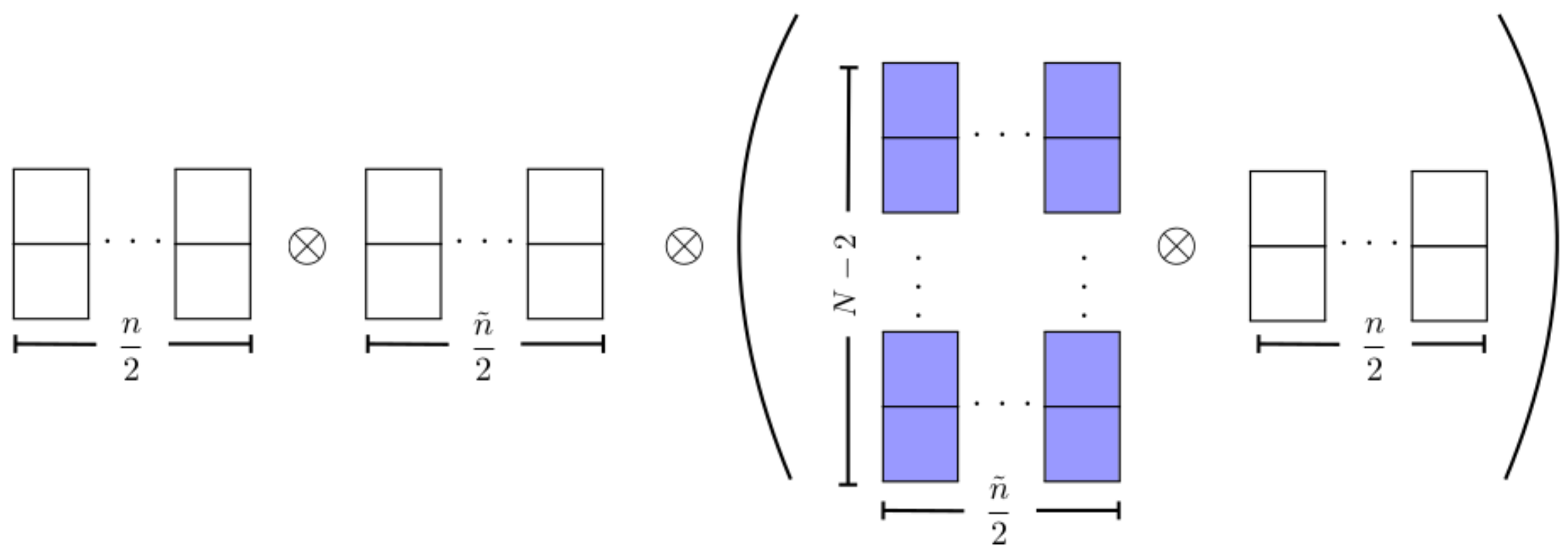} \end{array} \label{eq:decompa} \\
&=& \begin{array}{c}  \includegraphics[width=10cm]{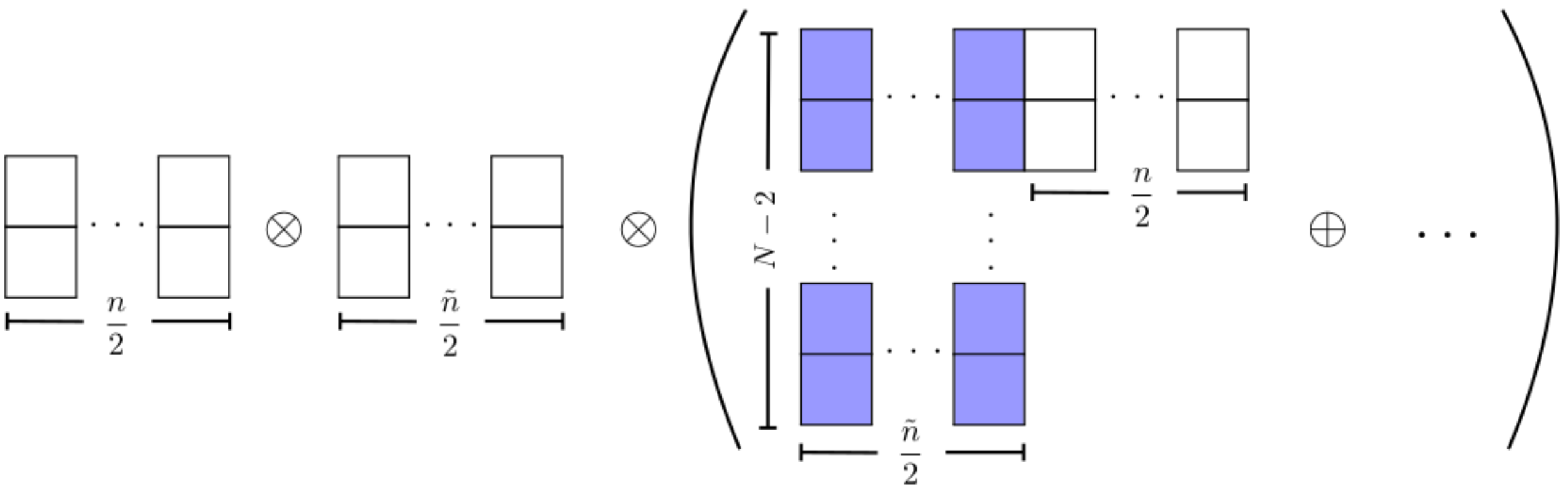} \end{array} \label{eq:decompb} \,.
\ee
In the last equality, the $U(N)$ tensor decomposition is indicated, displaying only the leading term; this
term coincides with the Young diagram in Fig.~\ref{fig:glue} and renders the polynomial harmonic, which we prove at the end of
this section.
This term is leading in the sense that it is the only $U(N)$ representation in the decomposition that does not
contain an overall factor of momentum, $P$, and thus the only primary operator/ harmonic mode in the decomposion.
We now turn to proving this.

The familiar diagrammatic `box placing' rules for carrying out tensor decompositions with Young diagrams (Littlewood-Richardson rules, again,
see {\it e.g.}~\cite{Georgi:1982jb}) can be applied to the product in  eq.~\eqref{eq:decompa}. The leading term appearing in
eq.~\eqref{eq:decompb} is in fact the simplest representation obtained using these rules---no white boxes have been shifted around, and the Young tableaux have been simply stuck together. 

What of the other `$\ldots$' terms  in eq.~\eqref{eq:decompb}? The box placing rules specify that we end up with a Young diagram that has either one or two white boxes at the bottom of a blue box column of length $N-2$. 
For the case of one white box under a column of $N-2$ blue boxes, we can factor a term in the resulting diagram that has the form
\be
&~&[j_1 j_2] \ep^{j_1j_2 k_1 .. k_{N-2}} \langle i_1 | + (\text{anti-sym in }k_1,\ldots,k_{N-2},i_1) \nonumber \,.
\ee
By the antisymmetry, the indices $k_1,\ldots,k_{N-2},i_1$ must be distinct choices of $1\ldots N$ (otherwise the anti-symmetrisation sets this factor to zero); without loss of generality, we consider the choice $1,\ldots,N-1$. Each cyclicly related set of terms in the above anti-sym is proportional (by a sign) to
\be
\sum_{k=1}^{N-1}[ N k ] \langle k | =   [N | P -  [N N] \langle N| = [N | P \,,
\label{eq:proof1}
\ee
using $P=\sum_{k=1}^N |k] \langle k|$ and $[ N N]=0$. Eq.~\eqref{eq:proof1}, as promised, contains a factor of total momentum, $P$, and  thus the operator is a descendent.

For the case of two white boxes under a column of $N-2$ blue boxes, one proceeds similarly: first, we  factor a term 
\be
&~&[j_1 j_2] \ep^{j_1j_2 k_1 .. k_{N-2}} \langle i_1 | \langle i_2 | + (\text{anti-sym in }k_1,..,k_{N-2},i_1, i_2) \nonumber \,.
\ee
(The spinors $ \langle i_1 |$ and $ \langle i_2 |$ could be contracted, $\langle i_1i_2\rangle$; the below arguments are valid in this case too.) The indices $k_1,\ldots,k_{N-2},i_1,i_2$ are anti-symmeterised permutations of the set $1..N$. Evidently, for any fixed value of $i_2$, one can factor out $P$ as per eq.~\eqref{eq:proof1}; in fact, one can easily show that in summing over the other values of $i_2$, a factor of $P^2$ can be pulled out overall.

This shows that the additional $U(N)$ representations are descendents, because they have the overall factor of $P$. 
We will return to a  proof that the leading Young diagram eq.~\eqref{eq:decompb} is annihilated by $K$ very shortly, showing that it is primary, after the introduction 
of semi-standard Young tableaux.

\subsection{States from semi-standard Young tableaux}

For a given Young diagram, one can construct the states of the corresponding $U(N)$ representation
using semi-standard Young tableau (SSYT), which we will see provides the labelling of the little group scaling. We recall that a SSYT is a filling
of the boxes of a Young diagram with the numbers 1 through $N$ (repeated use of a number is allowed) subject to the following rules:
\begin{itemize}
\item The numbers along the rows must weakly increase ({\it i.e.} reading from left to right each subsequent number
must be greater than or equal to the previous one) 
\item The numbers down the columns must strongly increase ({\it i.e.} reading from top to bottom each subsequent number
must be greater than the previous one) 
\end{itemize}

The number of valid SSYT is equal to the dimension of the $U(N)$ representation. For example, for the eight-dimensional 
adjoint representation of $U(3)$ we find eight SSYT fillings:
\be
\begin{ytableau} 
   1 &  1  \cr 
   2
\end{ytableau}~~~
\begin{ytableau} 
 1 &  1 \cr 
   3
\end{ytableau}~~~
\begin{ytableau} 
   1 &  2  \cr 
   2
\end{ytableau}~~~
\begin{ytableau} 
   1 &  2  \cr 
   3
\end{ytableau}~~~
\begin{ytableau} 
 1 &  3  \cr 
   2
\end{ytableau}~~~
\begin{ytableau} 
   1 &  3  \cr 
   3
\end{ytableau}~~~
\begin{ytableau} 
   2 & 2  \cr 
   3
\end{ytableau}~~~
\begin{ytableau} 
   2 &  3  \cr 
   3
\end{ytableau} \,. \nonumber
\ee
For a given SSYT of the Young diagram in Fig.~\ref{fig:glue}, one easily constructs the basis polynomial in $\la$ and $\lt$ 
using the diagram symmeterisation rules (sym on rows, anti-sym on columns). It is then straightforward to
read off the field content by the little group scaling for each particle; equivalently these are the eigenvalues of the $U(1)^N \subset U(N)$ generators. 
Note that the little group scaling of pairs of $\la^i$, $\lt_i$ cancel; for each such pair one should count a derivative to the 
field content of the harmonic/operator {\it i.e.} $\lambda^i_a\lt_{i \ad}=p^i_{a\ad}$ is the momentum of the $i$th particle (a derivative acting on the field for the $i$th particle). While each term in the polynomial must scale the same way under the 
little group overall,  the pairs of $\la^i$, $\lt_i$ could appear (and do appear) for different particle numbers $i$ in different terms.

We point out that the SSYT fillings will separately construct harmonics for all possible spins of each external state. For example, harmonics corresponding to each of the operators $F_{L\,1}F_{L\,2} \phi_3$, $F_{L\,1}\phi_{2} F_{L_3}$, and $\phi_1F_{L\,2}F_{L\,3}$ will be included separately. However, it is clear that these operators are of exactly the same form and can be 
related to each other with a simple particle index permutation. We emphasise we are dealing with all-distinguishable
particles, and that such a permutation is {\it between} particle species; it is not the (anti)-symmeterisation necessary when to describe indistinguishable particles. We can define a set of {\it reduced SSYT} which mods out such permutations between particle species with a simple ordering
rule:
\be
\text{order on SSYT filling: } \text{\#1s}\ge \text{\#2s}\ge\ldots \ge \text{\#$N$s} \,. \nonumber
\ee
That this is true is proven in the appendix.

As promised, we now return to the proof that all states of the representation shown in Fig.~\ref{fig:glue} are annihilated
by $K=-\sum\partial\widetilde{\partial}$. Consider the highest weight state, corresponding to the filling of all the boxes in the first row with 1s, all those
in the second row with 2s, {\it etc. }. Such a state is trivially annihilated by $K$: it consists only of polynomials in the four variables $\la_1$, $\la_2$, $\lt_{N-1}$ and $\lt_{N}$. The rest of the proof follows by group theory: since $K$ is a $U(N)$ singlet, its action commutes with the action of the $U(N)$ raising and lowering operators, and as such annihilates all the
states in the representation.

We conclude this section with a discussion on the orthogonality of the harmonics  constructed via the Young tableaux of Fig.~\ref{fig:glue}, under the phase space measure of eq.~\eqref{eq:phsp1}.\footnote{For an explicit formulation of phase space in terms of Grassmannian variables, see~\cite{Cox:2018wce}.} First, operators at different $N$ are orthogonal due to the Fock space structure of the  Hilbert space. Given the $U(N)$ symmetry of phase space, it is also clear that $U(N)$ representations with different  $n, \nt$ are automatically orthogonal. What of the states within each representation? The integral over the little group for each individual particle ensures that states with different eigenvalues
of the torus $U(1)^N$ are automatically orthogonal as well. In general, however, there exist degenerate subspaces where more than one operator has equal little group eigenvalues (the SSYT are permutations of each other). In such cases, state orthogonality is not guaranteed; we postpone discussion of this point (and details of normalisation with respect to the  phase space volume) to a future detailed, systematic study of the harmonics.

\section{EFT Spectra at low mass dimension}
\label{sec:lowmass}

It is instructive to work through the construction of harmonics/operators at low values of $n$ and $\nt$ {\it i.e.} at low mass
dimension, $\Delta$. In the following, we work through examples that suffice to construct an EFT basis up to mass dimension six. 

The formalism above provides a recipe to perform the construction:
\begin{enumerate}
\item Write down the Young diagram corresponding to the choice of $n$ and $\nt$, as shown in Fig.~\ref{fig:glue}.
\item Write down all  semi-standard Young tableau (SSYT) fillings to construct the $U(N)$ states.\footnote{Or any other method of constructing the states, {\it e.g.} start with the the highest weight state and apply lowering operators.}
\end{enumerate}
The operators we construct are summarised in Tables~\ref{tab:20},~\ref{tab:Neq3},~\ref{tab:Neq4}. 

We will highlight the special features of this conformal basis as we come across them. Of particular importance are the structure of the harmonics when annihilation by $K$ is  non-trivial. Such a case happens  when the corresponding operator involves derivatives, 
which is also where IBP relations come into play; these operators are necessarily non-holomorphic. Another
feature is the grouping of harmonics/operators with differing field content as states of the same $U(N)$ representation.

Below we normalise the Young tableaux permutations with a factor $1/k$, 
\be
k = \prod_{i\in \text{rows}} \prod_{j\in \text{columns}}  p_{i}! \,q_j!  \,,
\ee
where $p_i$ is the number of boxes in the $i$th row, and $q_j$ is the number of boxes in the $j$th column of the tableaux.

 \begin{table}

\begin{tabular}{c  c  }
\rule[-5ex]{0pt}{4ex}
~~ { 
 \begin{ytableau} 
1\cr 2 \cr
\end{ytableau}} ~~
&
~~ {
$\begin{array}{c}
\begin{ytableau} 
*(blue!40) 1  \cr *(blue!40) 2  \cr 
\end{ytableau}\\
\cdot\\\cdot\\
\begin{ytableau} 
 *(blue!40)   {\scriptscriptstyle N-2}\cr
\end{ytableau}
\end{array}$}
~~
 \\  
 $(2,0) $ & $(0,2) $ 
  \\  \hline
$\phi^{N-2} \psi_L^2$ & $\phi^{N-2} \psi_R^2$  \\
\end{tabular}

\caption{Reduced SSYT for Lorentz scalar operators of the form $(n,\nt)=(2,0), (0,2)$ for all $N\ge 3$.}
\label{tab:20}
\end{table}

\subsection{Harmonics of type $(n,\nt)=(2,0)$, $(0,2)$ }
We begin with harmonics for which $(n,\nt)=(2,0)$, $(0,2)$. These are the  simplest (non-trivial) harmonics, and we
consider them for all $N\ge 2$. The relevant reduced SSYT are displayed in Tab.~\ref{tab:20}. They correspond
to operators of field content $\phi^{N-2}\psi_L^2$ and $\phi^{N-2}\psi_R^2$, respectively. We re-emphasise that
we consider distinguishable particles at this point; the particle index is suppressed in the Table, but we indicate
it explicitly in the following construction:

\noindent$\phi_3\ldots\phi_N \psi_{L\,1} \psi_{L\,2}:$
\be
\begin{ytableau} 
1  \cr 2 \cr 
 \end{ytableau} ~= \frac{1}{2!}(\lambda^{1\,\al} \lambda^2_\al - \lambda^{2\,\al} \lambda^1_\al) = \langle 12\rangle  \,.
\label{eq:tab1a}
 \ee 
\noindent$ \phi_1\ldots\phi_{N-2} \psi_{R\,N-1} \psi_{R\,N}:$
 \begin{subequations}
 \be
\begin{array}{c}
\begin{ytableau} 
*(blue!40) 1  \cr *(blue!40) 2  \cr 
\end{ytableau}\\
\cdot\\\cdot\\
\begin{ytableau} 
 *(blue!40)   {\scriptscriptstyle N-2}\cr
\end{ytableau}
\end{array} ~&=&\frac{1}{(N-2)!} \lt_{j_1\,\ad}\lt_{j_2}^\ad(\ep^{ j_1j_2 1..N-2} + \text{anti-sym. in $1..N-2$}) \,, \label{eq:tab1bA} \\
&=& [N-1\,N] \,, 
\label{eq:tab1bB}
\ee
\end{subequations}
 where in eq.\eqref{eq:tab1bA} summation over $j_1$ and $j_2$ is implied. Putting back in the flavour permutations,
 there are $N(N-1)/2$ SSYT obtained from each of the reduced ones in eqs.~\eqref{eq:tab1a},~\eqref{eq:tab1bB}. 
 Note that the operators are conjugate to each other {\it i.e.}  $L\leftrightarrow R$ in all fields, and are
 thus related by switching $\la\leftrightarrow\lt$, or $\langle\,\rangle\leftrightarrow[\,]$.

\begin{table}
\begin{tabular}{ c c  c c   }
\rule[0ex]{0pt}{4ex} 
~~ { \begin{ytableau} 
1 & 1 \cr 2 & 2 \cr 
\end{ytableau}} ~~
&
~~ { \begin{ytableau} 
1 & 1 \cr 2 & 3 \cr 
\end{ytableau}} ~~
&
~~ { \begin{ytableau} 
*(blue!40)  1 & *(blue!40)  1 \cr 
\end{ytableau}} ~~
&
~~ { \begin{ytableau} 
*(blue!40)  1 & *(blue!40)  2 \cr 
\end{ytableau}} ~~
 \\  
   $(4,0) $  &   $(4,0)$  &  $(0,4) $  &   $(0,4) $    
 \\  
 $\phi F_L^2$ & $F_L \psi_L^2$ & $\phi F_R^2$ & $F_R \psi_R^2$ \\ \hline
 \end{tabular}
\begin{tabular}{ c c c }
\rule[1ex]{0pt}{3ex}  ~~ { \begin{ytableau} 
1 & 1 & 1 \cr 2 & 2  & 2\cr 
\end{ytableau}} ~~
&
~~ { \begin{ytableau} 
1 & 1 & 1 \cr 2 & 2 & 3 \cr 
\end{ytableau}} ~~&
~~ { \begin{ytableau} 
1 & 1 & 2 \cr 2 & 3 & 3 \cr 
\end{ytableau}} ~~
 \\  
   $(6,0) $  &   $(6,0)$  & $(6,0)$ 
 \\  
 $\phi \,\xi_L^2$ & $\xi_L F_L \psi_L$ & $ F_L^3 $ \\ \hline
\end{tabular}
\begin{tabular}{  c c  c }
\rule[1ex]{0pt}{3ex} 
~~ { \begin{ytableau} 
*(blue!40)  1 & *(blue!40)  1 & *(blue!40)  1 \cr 
\end{ytableau}} ~~
&
~~ { \begin{ytableau} 
*(blue!40)  1 & *(blue!40)  1 & *(blue!40)  2 \cr 
\end{ytableau}} ~~
&
~~ { \begin{ytableau} 
*(blue!40)  1 & *(blue!40)  2 & *(blue!40)  3 \cr 
\end{ytableau}} ~~
 \\  
   $(0,6) $  &   $(0,6) $     &   $(0,6) $    
 \\  
  $\phi\,\xi_R^2$ & $\xi_R F_R \psi_R$ & $ F_R^3 $\\ \hline
\end{tabular}
\caption{Reduced SSYT for Lorentz scalar operators with $N=3$, for low values of $n$, $\nt$. $\xi$ denotes a spin $3/2$ field.}
\label{tab:Neq3}
\end{table}

\subsection{Low `frequency' harmonics for $N=3$}

Next, we fix the number of particles in the harmonic to be $N=3$, and consider harmonics
of low $n$ and $\nt$. The case $N=3$ is special, as the construction given in Fig.~\ref{fig:glue} does not
produce a valid Young tableau when both $n$ and $\nt$ are non-zero. This reflects the fact that all
of the Lorentz scalar harmonics/operators for $N=3$ are holomorphic (or anti-holomorphic).  In Table~\ref{tab:Neq3} we consider
the cases $(n,\nt) = (4,0), (0,4),(6,0),(0,6)$, and show the reduced SSYT. 
Note how harmonics with different field content are grouped into the same $U(N)$ representation; for example
the harmonics $\phi F_L^2$ and $F_L \psi_L^2$ both appear as states in the $(4,0)$ representation.

The left-handed holomorphic $N=3$ operators in Tab.~\ref{tab:Neq3} are constructed as follows.

\noindent$\phi_3 F_{L\,1} F_{L\,2}:$
\be
\begin{array}{c}
\begin{ytableau} 
1 & 1 \cr 2 & 2 \cr 
\end{ytableau} \end{array} ~&=& \frac{1}{(2!)^4}(\lambda^{1\,\al} \lambda^2_\al \lambda^{1\,b} \lambda^2_b
 + \begin{array}{c}\text{tab. perms }\end{array} )  \,, \nonumber \\
 &=& \frac{2^2}{(2!)^4}(\lambda^{1\,\al} \lambda^2_\al \lambda^{1\,b} \lambda^2_b -\lambda^{2\,\al} \lambda^1_\al \lambda^{1\,b} \lambda^2_b -\lambda^{1\,\al} \lambda^2_\al \lambda^{2\,b} \lambda^1_b +\lambda^{2\,\al} \lambda^1_\al \lambda^{2\,b} \lambda^1_b ) \nonumber \,, \\
 &=& \langle 12 \rangle^2\,.
 \label{eq:tab2a}
 \ee
 \noindent$F_{L\,1} \psi_{L\,2}\psi_{L\,3}:$
\be
\begin{array}{c}
\begin{ytableau} 
1 & 1 \cr 2 & 3 \cr 
\end{ytableau} \end{array} ~ &=& \frac{2^2}{(2!)^4}(\lambda^{1\,\al} \lambda^2_\al \lambda^{1\,b} \lambda^3_b -\lambda^{2\,\al} \lambda^1_\al \lambda^{1\,b} \lambda^3_b -\lambda^{1\,\al} \lambda^2_\al \lambda^{3\,b} \lambda^1_b +\lambda^{2\,\al} \lambda^1_\al \lambda^{3\,b} \lambda^1_b ) \,, \nonumber \\
 &=& \langle 12 \rangle\langle 13 \rangle \,.
 \label{eq:tab2b}
 \ee
  \noindent $\phi_3 \xi_{L\,1}\xi_{L\,2}:$
\be
\begin{array}{c}
 \begin{ytableau} 
1 & 1 & 1 \cr 2 & 2  & 2\cr 
\end{ytableau} \end{array} ~ &=& \frac{1}{(3!)^2(2!)^3}(\lambda^{1\,\al} \lambda^2_\al \lambda^{1\,b} \lambda^2_b\lambda^{1\,c} \lambda^2_c + \begin{array}{c}\text{tab. perms }\end{array}) \,,\nonumber \\
 &=& \langle 12 \rangle^3 \,.
 \label{eq:tab2c}
 \ee
\noindent $\xi_{L\,1} F_{L\,2} \psi_{L\,3}:$
\be
\begin{array}{c}
 \begin{ytableau} 
1 & 1 & 1 \cr 2 & 2  & 3\cr 
\end{ytableau} \end{array} ~ &=& \frac{1}{(3!)^2(2!)^3}(\lambda^{1\,\al} \lambda^2_\al \lambda^{1\,b} \lambda^2_b\lambda^{1\,c} \lambda^3_c + \begin{array}{c}\text{tab. perms }\end{array}) \,, \nonumber \\
 &=& \langle 12 \rangle^2\langle 13 \rangle \,.
 \label{eq:tab2d}
 \ee
   \noindent $F_{L\,1}F_{L\,2}F_{L\,3}:$
\be
\begin{array}{c}
 \begin{ytableau} 
1 & 1 & 2 \cr 2 & 3  & 3\cr 
\end{ytableau} \end{array} ~ &=& \frac{1}{(3!)^2(2!)^3}(\lambda^{1\,\al} \lambda^2_\al \lambda^{1\,b} \lambda^3_b\lambda^{2\,c} \lambda^3_c + \begin{array}{c}\text{tab. perms }\end{array}) \,, \nonumber \\
 &=& \langle 12 \rangle\langle 13 \rangle\langle 23 \rangle \,.
 \label{eq:tab2e}
 \ee
 In the above we made use of the formula $\la^{i\,\al}\la^j_\al =-\la^{i}_{\al}\la^{j\,\al}$. The remaining
 right-handed holomorphic operators in Tab.~\ref{tab:Neq3} can be obtained (up to flavour permutations)
 by exchanging $\la\to \lt$ {\it i.e.} $\langle i j\rangle\to[ij]$. However, we work out one case from the tableaux explicitly,
 for illustrative purposes,
 
\noindent $\psi_{R\,1}\psi_{R\,2} F_{R\,3}:$
\be
\begin{array}{c}
 \begin{ytableau} 
*(blue!40)  1 & *(blue!40)  2 \cr 
\end{ytableau} \end{array} ~ &=&\frac{1}{2!}\lt_{j_1\, \ad}\lt_{j_2}^{\ad}\lt_{j_3\,\dot{b}}\lt_{j_4}^{\dot{b}} ( \ep^{j_1 j_2 1}  \ep^{j_3 j_4 2}+\ep^{j_1 j_2 2}  \ep^{j_3 j_4 1} ) \,,
 \nonumber \\
 &=& [13 ][32] \,,
 \label{eq:tab2f}
 \ee
 where in the above summation over $j_1,j_2,j_3,j_4$ is implied. This is indeed the conjugate of (a flavour permutation of) eq.~\eqref{eq:tab2b}.

\begin{table}
\begin{tabular}{c c c c}
\rule[-5ex]{0pt}{4ex} &
~~~\begin{ytableau} 
~&~ \cr ~ &~ \cr
\end{ytableau}~~~
&~~~ \begin{ytableau} 
*(blue!40) & ~ \cr *(blue!40)   & ~ \cr
\end{ytableau}~~~ 
&~~~ \begin{ytableau} 
*(blue!40) & *(blue!40)  \cr *(blue!40)   & *(blue!40)  \cr
\end{ytableau}~~~\\  
~ & $(4,0)$ & $(2,2)$ & $(0,4)$ \\\hline
\rule{0pt}{4ex}~~{\tiny \begin{ytableau} 
1 & 1 \cr 2 &2 \cr
\end{ytableau}}~~ &  $\phi^2 F_L^2$ & $\psi_L^2\psi_R^2$& $\phi^2 F_R^2$
\\
\rule{0pt}{4ex}~~{\tiny \begin{ytableau} 
1 & 1 \cr 2 & 3 \cr
\end{ytableau}}~~ &  $\phi F_L \psi_L^2 $ & $\phi^2 \psi_L \psi_R D$& $\phi F_R \psi_R^2$
\\
\rule{0pt}{4ex}~~{\tiny \begin{ytableau} 
1 & 2 \cr 3 &4  \cr
\end{ytableau}}~~ &  $\psi_L^4$ & $\phi^4D^2$& $\psi_R^4$
\\
\rule{0pt}{4ex}~~{\tiny \begin{ytableau} 
1 & 3 \cr 2 &4 \cr
\end{ytableau}}~~ &  $\psi_L^4$ & $\phi^4D^2$& $\psi_R^4$
\\
\end{tabular}
\caption{Reduced SSYT  for Lorentz scalar operators with with $N=4$, for low values of $n$, $\nt$.}
\label{tab:Neq4}
\end{table}

\subsection{Low `frequency' harmonics for $N=4$}

As a last class of examples, we consider harmonics involving $N=4$ fields; reduced SSYT for $(n,\nt)=(4,0),(2,2),(0,4)$ are shown in Tab.~\eqref{tab:Neq4}. There are two new features evident in the Table that were not present for the 
cases considered above. The first feature is that now non-holomorphic harmonics appear (the middle column). We will
comment on the detailed form of these operators below. The second feature is that there are distinct harmonics with 
the same field content: two copies of the harmonic $\psi_L^4/\psi_R^4$ appear in the $(4,0)/(0,4)$ $U(N)$ representation, and
two copies of $\phi^4 D^2$ appear in the $(2,2)$ representation. These operators are independent, so it is important
that they are both included; the rules for constructing the reduced SSYT  ensure this happens.

The left-handed holomorphic ones are constructed as follows (the first two are identical to the operators in eq.~\eqref{eq:tab2a} and eq.~\eqref{eq:tab2b}, respectively, differing only by the addition of an extra $\phi$ field).
 
\noindent$\phi_3\phi_4 F_{L\,1} F_{L\,2}:$
\be
\begin{array}{c}
\begin{ytableau} 
1 & 1 \cr 2 & 2 \cr 
\end{ytableau} \end{array} ~&=& \langle 12 \rangle^2 \,.
 \label{eq:tab3a}
 \ee
 \noindent$\phi_4 F_{L\,1} \psi_{L\,2}\psi_{L\,3}:$
\be
\begin{array}{c}
\begin{ytableau} 
1 & 1 \cr 2 & 3 \cr 
\end{ytableau} \end{array} ~&=& \langle 12 \rangle  \langle 13 \rangle   \,.
 \label{eq:tab3b}
 \ee
\noindent$ \psi_{L\,1}\psi_{L\,2}\psi_{L\,3}\psi_{L\,4}:$
\be
\begin{array}{c}
\begin{ytableau} 
1 & 2 \cr 3 & 4 \cr 
\end{ytableau} \end{array} ~&=&\frac{1}{(2!)^4}(\lambda^{1\,\al} \lambda^3_\al \lambda^{2\,b} \lambda^4_b
 + \begin{array}{c}\text{tab. perms }\end{array} ) \,,\nonumber \\
 &=& \frac{1}{2} ( \langle 14\rangle \langle 23\rangle+ \langle 13\rangle \langle 24\rangle ) \,.
 \label{eq:tab3c}
 \ee
 \noindent$ \psi_{L\,1}\psi_{L\,2}\psi_{L\,3}\psi_{L\,4}:$ 
 \be
\begin{array}{c}
\begin{ytableau} 
1 & 3 \cr 2 & 4 \cr 
\end{ytableau} \end{array} ~&=&\frac{1}{(2!)^4}(\lambda^{1\,\al} \lambda^2_\al \lambda^{3\,b} \lambda^4_b
 + \begin{array}{c}\text{tab. perms }\end{array} )  \,, \nonumber \\
 &=& \frac{1}{2} ( -\langle 14\rangle \langle 23\rangle+ \langle 12\rangle \langle 34\rangle ) \,.
 \label{eq:tab3d}
 \ee
The right-handed holomorphic harmonics in Tab.~\ref{tab:Neq4} are obtained via conjugation of eqs.~\eqref{eq:tab3a}-\eqref{eq:tab3d}, and so we do not present their construction explicitly.
 
Turning finally to  the non-holomorphic harmonics, we have,

 \noindent$\psi_{L\,1} \psi_{L\,2}\psi_{R\,3} \psi_{R\,4}:$
\be
\begin{array}{c}
\begin{ytableau} 
*(blue!40) 1 & 1 \cr *(blue!40)  2 & 2 \cr
\end{ytableau} \end{array} ~&=& \frac{1}{(2!)^4}\lt_{j_1\,\ad}\lt_{j_2}^{\ad}( \ep^{ j_1 j_21 2} \la^{1 \,\al} \la^{2}_\al  + \begin{array}{c}\text{tab. perms }\end{array} )  \,, \nonumber \\
&=&\frac{2^2}{(2!)^4}\lt_{j_1\,\ad}\lt_{j_2}^{\ad} (\ep^{j_1 j_2 1 2} \la^{1 \,\al} \la^{2}_\al -   \ep^{j_1 j_2 21} \la^{1 \,\al} \la^{2}_\al  - \ep^{ j_1 j_21 2} \la^{2 \,\al} \la^{1}_\al  + \ep^{ j_1 j_221} \la^{2\,\al} \la^{1}_\al ) \,, \nonumber \\
&=& \langle 12\rangle [34] \,.
 \label{eq:tab3e}
 \ee
  \noindent$\phi_{2} \phi_{3} \psi_{L\,1}\psi_{R\,4} D :$ 
\be
\begin{array}{c}
\begin{ytableau} 
*(blue!40) 1 & 1 \cr *(blue!40)  2 & 3 \cr
\end{ytableau} \end{array} ~&=& \frac{1}{(2!)^4}\lt_{j_1\,\ad}\lt_{j_2}^{\ad}( \ep^{ j_1 j_21 2} \la^{1 \,\al} \la^{3}_\al  + \begin{array}{c}\text{tab. perms }\end{array} )  \,, \nonumber \\
&=&\frac{2}{(2!)^4}\lt_{j_1\,\ad}\lt_{j_2}^{\ad} (\ep^{ j_1 j_2 1 2} \la^{1 \,\al} \la^{3}_\al +  \ep^{j_1 j_2 13} \la^{1 \,\al} \la^{2}_\al  + \begin{array}{c}\text{tab. anti-syms}\end{array}  ) \,, \nonumber \\
&=& \frac{1}{2} (\langle 13\rangle  [34]-\langle 12\rangle  [24] ) \,.
 \label{eq:tab3f}
 \ee
   \noindent$\phi_1 \phi_{2}\phi_{3} \phi_{4} D^2 :$ 
\be
\begin{array}{c}
\begin{ytableau} 
*(blue!40) 1 & 2 \cr *(blue!40)  3 & 4 \cr
\end{ytableau} \end{array} ~&=& \frac{1}{(2!)^4}\lt_{j_1\,\ad}\lt_{j_2}^{\ad}( \ep^{ j_1 j_2 13} \la^{2 \,\al} \la^{4}_\al  + \begin{array}{c}\text{tab. perms }\end{array} ) \,, \nonumber \\
&=&\frac{1}{(2!)^4}\lt_{j_1\,\ad}\lt_{j_2}^{\ad} (\ep^{ j_1 j_213} \la^{2 \,\al} \la^{4}_\al  + \ep^{ j_1 j_223} \la^{1 \,\al} \la^{4}_\al \nonumber \\
&~&~~+\ep^{ j_1 j_214} \la^{2 \,\al} \la^{3}_\al +\ep^{ j_1j_224} \la^{1 \,\al} \la^{3}_\al + \begin{array}{c}\text{tab. anti-syms}\end{array}  )  \,, \nonumber \\
&=& \frac{1}{4} (\langle 24\rangle  [42]-\langle 14\rangle  [41]-\langle 23\rangle  [32]+\langle 13\rangle  [31] ) \,.
 \label{eq:tab3g}
 \ee
    \noindent$\phi_1 \phi_{2}\phi_{3} \phi_{4} D^2 :$ 
\be
\begin{array}{c}
\begin{ytableau} 
*(blue!40) 1 & 3 \cr *(blue!40)  2 & 4 \cr
\end{ytableau} \end{array} ~&=& \frac{1}{(2!)^4}\lt_{j_1\,\ad}\lt_{j_2}^{\ad}( \ep^{ j_1 j_212} \la^{3 \,\al} \la^{4}_\al  + \begin{array}{c}\text{tab. perms }\end{array} )  \,, \nonumber \\
&=& \frac{1}{4} (\langle 34\rangle  [43]-\langle 14\rangle  [41]-\langle 23\rangle  [32]+\langle 12\rangle   [21] ) \,.
 \label{eq:tab3h} 
 \ee
 The last three of these have non-trivial annihilation by $K$.
 For example, the harmonic in eq.~\eqref{eq:tab3f} with operator 
content $\phi^2 \psi_L \psi_R D$,
\be
\sum_{i=1}^N \frac{\partial}{\partial \lt_i^{\ad}}  \frac{\partial}{\partial \la^{i \,\al}} \left(\frac{1}{2} (\langle 13\rangle  [34]-\langle 12\rangle  [24] ) \right) = \frac{1}{2}\bigg( \la^{1\,\al} \lt_4^\ad  -  \la^{1\,\al} \lt_4^\ad  \bigg) =0\,.
\label{eq:specex}
\ee
Using momentum conservation, one could rewrite the operator eq.~\eqref{eq:tab3f} as another equally valid operator basis element, {\it e.g.} simply $\langle 13\rangle  [34]$ or $\langle 12\rangle  [24]$, but it is only the combination $\propto (\langle 13\rangle  [34]-\langle 12\rangle  [24] )$ that that is a conformal primary and is annihilated by $K$ as in eq.~\eqref{eq:specex}; it is in this sense that the harmonics form a privileged basis.

\section{Discussion}
\label{sec:discuss}

The general construction above applies to the distinguishable particles case. To take into account exchange symmetry one must (anti-)symmeterise over the  identical (fermionic) bosonic fields in an operator. The particle index can also be interpreted as a gauge or other
symmetry index; further bookkeeping is required here too. The kinematic construction detailed here is a necessary
first step (and the above considerations can be easily applied by hand, if not entirely systematically at present).

To the EFTer, the systematic nature of the construction 
is clearly appealing. The automatic orthogonality of (the majority of operators) at different $N$ and with different $U(1)^N$ eigenvalue of basis elements also has utility: converting from a UV Lagrangian/other EFT parameterisation is then simple, via a projection $\int d\Phi_N Y^* \mathcal{L}_{other}$. It will be useful to further study orthogonality in the degenerate eigenvalue case. It would also be interesting to explore how this `mathematically singled out' basis fares in phenomenological applications.

There is deep structure in the operator basis which should be explored further. One of the interesting features is the mixing of different particle species within the same harmonic ({\it e.g.} the columns in Tab.~\ref{tab:Neq4})---does this imply any relation
between different phenomenological observables? We note that these harmonic blocks are  the same grouping as the classes in the non-renormalisation theorems~\cite{Alonso:2014rga,Cheung:2015aba,Elias-Miro:2014eia}, and may shed further light on the structure of EFT anomalous dimension matrices/amplitude non-interference~\cite{Azatov:2016sqh} results.  Of further interest is whether the harmonic picture presented here  sheds further light or provides tools for studing positivity-type constraints on Wilson coefficients~\cite{Froissart,Gribov:1961ex,Adams:2006sv,Bellazzini:2017fep,deRham:2017zjm}; it would also be interesting to understand the connection between this natural basis and  natural bases for amplidutes {\it e.g.} partial waves.

\section*{Acknowledgements}
We thank Peter Cox,  Marc Riembau, and Francesco Riva for conversations, and Rodrigo Alonso and Peter Cox for comments on the draft.
BH is funded by the Swiss National Science Foundation under grant no. PP002-170578. TM is supported by the World Premier International Research Center Initiative (WPI), MEXT, Japan, and by JSPS KAKENHI Grant Number JP18K13533.

\appendix
\section{Reduced tableaux}
\label{sec:app}

When operators are related by simple index permutations between particle species that do not change the form of the operator, {\it e.g.}  $F_{L\,1}F_{L\,2} \phi_3$, $F_{L\,1}\phi_{2} F_{L_3}$, and $\phi_1F_{L\,2}F_{L\,3}$, we wish to define a rule so as to only consider one of them. A canonical choice is to keep only  operators in which the fields are helicity ordered, such that those of lower helicity are assigned lower particle indices. In the example above, this would be the operator $F_{L\,1}F_{L\,2} \phi_3$. (Right handed fields have positive helicity, so if we replace all instances of $L\to R$ in the above example, the canonical choice would be $\phi_1 F_{R\,2}F_{R\,3}$.)

 More precisely, an operator is {\it not} of this canonical form if the following is true: {\it there exists a pair of fields in the operator that have particle index $i$ and $j$ with $i<j$, but have helicities satisfying $h_i>h_j$}. After removing such operators, we call the remaining set reduced operators.
%Such an ordering  mods out index permutations between the set of all fields of helicity $h_{\text{set } 1}$ and the set of all fields of helicity $h_{\text{set } 2}$, for all $h_{\text{set } 1}$ and $h_{\text{set } 2}$  with  $h_{\text{set } 2}\ne h_{\text{set } 1}$.  That is, it mods out permutations between the indices of different particle species, which are trivial in the sense that they do not change the form the operator takes.
We will show that the SSYT corresponding to a reduced operator
 satisfies 
\be
\text{order on SSYT filling: } \text{\#1s}\ge \text{\#2s}\ge\ldots \ge \text{\#$N$s} \,. 
\label{eq:toprove}
\ee

Before turning to the proof, note that if $h_i=h_j$ there is no notion of a canonical order on $i$ or $j$ in defining a reduced operator. That is, the set of reduced operators includes operators related by non-trivial permutations of the indices between fields of equal helicity.  For illustratation, we take two examples from the text. First, consider the SSYT for $N=3$,
\be
\begin{array}{c}
\begin{ytableau} 
1 & 1 \cr 2 & 3 \cr 
\end{ytableau} \end{array}~~:~~ F_{L\,1} \psi_{L\,2}\psi_{L\,3} \,.
\ee
The corresponding operator has a single field of helicity $-1$ and two fields of helicity $-\frac{1}{2}$, and it is of reduced form. There are no non-trivial  index permutations between the two fermions. (The Young tableaux corresponding to this permutation is not  semi-standard---it would be the filling $((1,1),(3,2))$.)  However, there are index permutations between the sets of fields to create the operators $F_{L\,2} \psi_{L\,1}\psi_{L\,3}$ and $F_{L\,3} \psi_{L\,1}\psi_{L\,2}$, which correspond to SSYT fillings $((1,2),(2,3))$ and $((1,2),(3,3))$. These two operators are not reduced, and are discarded by the ordering rule above; indeed, the SSYT fillings do not satisfy eq.~\eqref{eq:toprove}.

For the second example consider the SSYTs for $N=4$,
\be
\begin{array}{c}
\begin{ytableau} 
1 & 2 \cr 3 & 4 \cr 
\end{ytableau} \end{array}~~\text{and}~~\begin{array}{c}
\begin{ytableau} 
1 & 3 \cr 2 & 4 \cr 
\end{ytableau} \end{array} \,,
\label{eq:example2}
\ee
both of which are operators with field content $\psi_{L\,1} \psi_{L\,2} \psi_{L\,3} \psi_{L\,4}$ and are related by non-trivial particle index permutations between particles of equal helicity. Both are included in the reduced set of operators.\footnote{This highlights an issue with how the  particle index permutations are implemented across the set of reduced operatorss. To illustrate this, denote the two operators from the example~\eqref{eq:example2} as $(\psi_{L\,1} \psi_{L\,2} \psi_{L\,3} \psi_{L\,4})^A$ and $(\psi_{L\,1} \psi_{L\,2} \psi_{L\,3} \psi_{L\,4})^B$. Now consider  two reduced
operators that exist in the $N=7$ ring: $(\psi_{L\,1} \psi_{L\,2} \psi_{L\,3} \psi_{L\,4})^A\phi_5 \phi_6\phi_7$ and $(\psi_{L\,1} \psi_{L\,2} \psi_{L\,3} \psi_{L\,4})^B\phi_5 \phi_6\phi_7$. These are related by index permutations to the non-reduced operators $(\psi_{L\,1} \psi_{L\,5} \psi_{L\,6} \psi_{L\,7})^A\phi_2 \phi_3\phi_4$ and $(\psi_{L\,1} \psi_{L\,5} \psi_{L\,6} \psi_{L\,7})^B\phi_2 \phi_3\phi_4$. It would be incorrect to perform the permutation to the two  operators differently, such that one could obtain {\it e.g.} $(\psi_{L\,1} \psi_{L\,5} \psi_{L\,6} \psi_{L\,7})^A\phi_2 \phi_3\phi_4$ and $(\psi_{L\,1} \psi_{L\,6} \psi_{L\,5} \psi_{L\,7})^B\phi_2 \phi_3\phi_4$, which are in fact identical operators.}

We now turn to proving that the statement  on SSYT in eq.~\eqref{eq:toprove} follows for an operator that is of reduced form. 
First, consider the holomorphic case.
Here,  each field of helicity $h_i$ necessitates $2|h_i|$ copies of $\lambda^i$ in the operator, which in turn necessitates $2|h_i|$ copies of the box $\begin{array}{c}\begin{ytableau} i \end{ytableau} \end{array}$ in the SSYT filling.  Since for a reduced operator $|h_1|\ge |h_2|\ge\ldots\ge |h_N|$, and all helicities $h_i \le0$  in the holomorphic case,  the condition eq.~\eqref{eq:toprove} immediately follows.

Next, consider the anti-holomorphic case, where all the helicities $h_i\ge0$. A field of helicity $h_i$ necessitates $2h_i$ copies of $\lt_i$ in the operator. Each Lorentz contracted pair of $\lt_{i \,\ad} \lt_j^\ad$ necessitates a column in the SSYT of $N-2$ blue boxes filled with the numbers $1$ to $N$, excluding $i$ and $j$.  Since $h_N\ge h_{N-1} \ge\ldots\ge h_1$, the number $N$ will be excluded in the SSYT more (or equal) times than the number $N-1$, which in turn will be excluded more (or equal) times than $N-2$, {\it etc.}, and again the condition eq.~\eqref{eq:toprove} follows.

For the non-holomorphic case, first let us assume that no derivatives are present in the operator. In this case, we split the particles into negatice helicity (to which we apply the same reasoning in the holomorphic case) and into positive helicity (to which we apply the non-holomorphic reasoning), and conclude again that the condition eq.~\eqref{eq:toprove} holds.

Finally we need to show that derivatives do not change the counting. A derivative implies a pair $\lt_i\, \la^i$  (no sum on $i$) in the operator. It is useful to consider the $\lt_i$ as contributing $N-1$ boxes to the SSYT filled with the numbers $1$ to $N$, excluding $i$; when it is contracted with a $\lt_j$, a box $j$ is further removed. The $\la^i$ in the pair contributes a (white) box $\begin{array}{c}\begin{ytableau} i \end{ytableau} \end{array}$. Thus we see that the contribution of  $\lt_i \,\la^i$ to the SSYT filling is to add a set of $N$ boxes that contains one copy each of the numbers $1$ to $N$. As such, it does not affect the condition eq.~\eqref{eq:toprove}.

\bibliographystyle{utphys}
\bibliography{bibliography}

\end{document}